\documentclass[doublecol]{epl2}
\usepackage{pdflscape}
\usepackage{amsmath}
\usepackage{amssymb}
\newcommand{\rL}{\rho_\Lambda}

\newcommand{\CC}{\Lambda}

\newcommand{\OL}{\Omega_{\Lambda}}

\newcommand{\rmm}{\rho_{m}}
\newcommand{\rdm}{\rho_{ dm}}
\newcommand{\rb}{\rho_{ b}}

\newcommand{\rmo}{\rho_{m 0}}

\newcommand{\rmr}{\rho_m}

\newcommand{\rD}{\rho_D}

\newcommand{\pL}{p_{\CC}}
\newcommand{\rLo}{\rho_{\CC 0}}
\newcommand{\pD}{p_D}
\newcommand{\wD}{w_D}

\newcommand{\be}{\begin{equation}}
\newcommand{\ee}{\end{equation}}
\def\beq{\begin{equation}}
\def\eeq{\end{equation}}
\def\ber{\begin{eqnarray}}
\def\eer{\end{eqnarray}}

\title{Dynamical dark energy  versus  $\CC=$const.  in light of observations}
\author{Joan Sol\`a Peracaula, Javier de Cruz P\'erez and Adri\`a G\'omez-Valent} %$^{a,b}$,}%$^{a}$}
\institute{%$^{a}$
Departament de F\'isica Qu\`antica i Astrof\'isica, and Institute of Cosmos Sciences, Universitat de Barcelona, \\ Av. Diagonal 647, E-08028 Barcelona, Catalonia, Spain}%\\

\abstract{After about two decades of the first observational papers confirming the  accelerated expansion of the universe, we are still facing the question whether the cause of it is a rigid cosmological constant $\CC$-term or a mildly evolving dynamical dark energy (DDE). While studies focusing mainly on CMB measurements do not perceive signs of physics beyond the $\CC$CDM, in this work we show that if we take a large string SNIa+BAO+$H(z)$+LSS+CMB of modern cosmological observations, in which not only the CMB but also a rich sample of large scale structure formation data are included, one can extract $\sim 3.3\sigma$ signs of DDE using a simple XCDM parameterization. These signs can be enhanced up to near $3.8\sigma$ in the context of the running vacuum model (RVM), in which the vacuum energy density is in interaction with dark matter.  Recently the RVM has been shown to provide an efficient and economical solution to the $\sigma_8$-tension, which is one of the intriguing phenomenological problems that has not been possible to solve within the $\CC$CDM so far. This fact contributes to strengthen the possibility that dynamical vacuum energy, or in general DDE, could be presently favored by the observations. }

\begin{document}

\maketitle

\section{Introduction}
\label{intro}

The cosmological constant,  $\CC$,  has been part of Einstein's field equations  for  101 years\,\cite{Einstein1917}, but we still ignore its physical meaning and its ultimate theoretical nature. In the meantime, we firmly observe from different sources of experimental evidence that the universe is in accelerated expansion\,\cite{SNIaRiess,SNIaPerl,Planck2015}, and the simplest hypothesis that can be made to explain it has been to assume that $\CC$ is  a nonvanishing and positive constant. This is tantamount to saying that one assumes that the vacuum energy density  $\rL=\CC/(8\pi G)$ ($G$ being Newton's gravitational coupling) remains constant throughout the expansion. This is the crucial point of view advocated by the concordance or $\CC$CDM model, and is essentially supported by the observations. However, the $\CC$-term harbors one of the most profound (and unresolved) theoretical enigmas of fundamental physics: the cosmological constant (CC) problem\,\cite{Weinberg,CCP}, namely the preposterous mismatch between the typical prediction for  $\CC$ in quantum field theory (QFT) -- e.g. in the standard model of particle physics --  and the measured value from cosmological observations. For this reason $\CC$ has been promoted into the multifarious concept of dark energy (DE)\,\cite{DEBook}. In performing this generalization the  new concept is  no longer a constant but a dynamical variable (slowly varying with the expansion).  For example, scalar field models have been proposed since long ago either to adjust dynamically the value of $\rL$ (e.g. the cosmon model\,\cite{PSW}) or to explain the coincidence problem with the notion of quintessence and the like \cite{PeeblesRatra88,Wetterich88,Caldwell98,ZlatevWangSteinhardt99,Amendola2000}, among many other alternatives, see\,\cite{CCP,DEBook}. {In particular, we have the modified gravity theories, such as the class of $f(\mathcal{R})$ models, which has proven capable of triggering the current acceleration of the universe and improve the cosmographic description, see e.g. \cite{Capozziello2014}.}

%%%%%%%%%%%%%%%%%%%%%%%%%%%%%%%%%%%%%%%%%%%%%%%%%%%%%%%%%%%%%%%%%
%%%%%%%%%%%%%%%%%%%%%%%%%%%%%%%%%%%%%%%%%%%%%%%%%%%%%%%%%%%%%%%%%

\begin{table*}
\begin{center}
\begin{scriptsize}
%\resizebox{1\textwidth}{!}{
\begin{tabular}{| c | c |c | c | c | c | c | c | c | c|}
\multicolumn{1}{c}{Model} &  \multicolumn{1}{c}{$h$} &  \multicolumn{1}{c}{$\omega_b= \Omega_b h^2$} & \multicolumn{1}{c}{{$n_s$}}  &  \multicolumn{1}{c}{$\Omega_m$}&  \multicolumn{1}{c}{$\nu(\times 10^{-3})$}  & \multicolumn{1}{c}{$w_0$} &
\multicolumn{1}{c}{$\chi^2_{\rm min}/dof$} & \multicolumn{1}{c}{$\ln A$} & \multicolumn{1}{c}{$\ln B$}\vspace{0.5mm}
\\\hline
$\CC$CDM & $0.692\pm 0.004$ & $0.02253\pm 0.00013$ &$0.976\pm 0.004$& $0.296\pm 0.004$ & - & -1 &  84.88/85 & - & -\\
\hline
XCDM  &  $0.672\pm 0.007$& $0.02262\pm 0.00014 $&$0.976\pm0.004$& $0.311\pm 0.007$& - & $-0.923\pm0.023$ &  74.08/84 & 4.28 & 3.16 \\
\hline
RVM  & $0.677\pm 0.005$& $0.02231\pm 0.00014$&$0.965\pm 0.004$& $0.303\pm 0.005$ & $1.58\pm 0.42$ & -1 & 69.72/84 & 6.46 & 5.34
\\
\hline
 $\Lambda$CDM$^{*}$& $0.687\pm 0.004$& $0.02246\pm 0.00013$&$0.971\pm 0.004$& $0.302\pm 0.005$ & - & -1  &  65.93/73 & - & -\\
\hline
  XCDM$^{*}$& $0.678\pm 0.008$& $0.02253\pm 0.00015$&$0.973\pm 0.004$& $0.308\pm 0.007$ & - & $-0.957\pm 0.031$ & 64.33/72 & -0.35 & -1.37 \\
\hline
RVM$^{*}$  & $0.680\pm 0.006$& $0.02235\pm 0.00015$&$0.966\pm 0.004$& $0.303\pm 0.005$ & $1.05\pm 0.60$ & -1 & 62.58/72 & 0.53 & -0.50 \\
\hline
\end{tabular}
% }
\end{scriptsize}
\end{center}
\caption{Best-fit values for  $\CC$CDM, XCDM and the running vacuum model (RVM), including the $\chi^2$-test and Akaike and Bayesian evidence criteria. These criteria favor the DDE options since $\ln A,\ln B>0$ for them. We use a rich and updated SNIa+BAO+$H(z)$+LSS+CMB data set, to wit:  $31$ effective points from the JLA sample of SNIa\,\cite{BetouleJLA}, $11$ from BAO\,\cite{Beutler2011,Ross2015,Kazin2014,GilMarin2017,Delubac2015,Aubourg2015}, $30$  from $H(z)$\,\cite{Zhang2014,Jimenez2003,Simon2005,Moresco2012,Moresco2016,Stern2010,Moresco2015}, $13$ from $f(z)\sigma_8(z)$ (LSS, mostly RSD) \cite{GilMarin2017,Beutler2012,Feix2015,Simpson2016,Blake2013,Blake2011BAO,Springob2016,Granett2015,Guzzo2008,SongandPercival2009}, and $4$ from CMB \cite{Huang2015}. See \, \cite{ApJ2017,PRD2017} for more details. Apart from the standard parameters ($h,\omega_b,n_s,\Omega_m$), the specific ones for each model are:  $w_0$ (XCDM) and $\nu $ (RVM). The fit contour lines for these models are shown in Fig.\, 1. The starred scenarios describe the fit results when the LSS data (i.e. the $f(z)\sigma_8(z)$ points from Fig. 2) are replaced with the $S_8$ value obtained from the analysis of the weak gravitational lensing data by KiDS-450 + 2dFLenS+BOSS\,\cite{Joudaki2017}, see text. The DDE signal for these scenarios is weaker. {The values of $h$ obtained for the various models lie in the lower range, i.e. are more resonant with those preferred by Planck \cite{Planck2015} rather than with the local value presented in \cite{Riess2016}. They are also more aligned with the estimates from recent model-independent analyses \cite{Capozziello2017,GomAmen2018}.}}
\label{tableFit1}
\end{table*}
%

%%%%%%%%%%%%%%%%%%%%%%%%%%%%%%%%%%%%%%%%%%%%%%%%%%%%%%%%%%%%%%%%%
%%%%%%%%%%%%%%%%%%%%%%%%%%%%%%%%%%%%%%%%%%%%%%%%%%%%%%%%%%%%%%%%%

Let us, however, remark that the problems of modern cosmology are not just of theoretical nature, such as the aforementioned CC problem.  That this is so is proven by the fact that other, more recent, observational pitfalls have been plaguing the  straightforward viability of the $\CC$CDM at the phenomenological level.  For instance, the so-called $H_0$\,\cite{Riess2016} and $\sigma_8$\,\cite{Macaulay2013,BasilakosNesseris2017} tensions are among the most prominent ones and demonstrate the existence of significant and persistent discrepancies of the concordance model prediction with the cosmological observations.  Altogether they seem to  indicate  that the idea of a strictly constant $\CC$ could be an oversimplification, even at the pure phenomenological level.  This should not be too surprising since the vacuum energy density of a dynamical universe in expansion may be better conceived as a dynamical quantity as well, and if so it should help to smooth out the current phenomenological conundrums. In short, whether $\CC$  is truly a constant or not is a matter that at this point should be settled empirically. This means to compare the $\CC$CDM ability to describe the bulk of the cosmological observations with other models in which the vacuum energy density $\rL$ is mildly evolving with time. There are recent hints that this could be the case, see, e.g., \cite{ApJL2015,ApJ2017} and  \cite{GongBoZhao2017}. Furthermore, some of these scenarios could offer a solution to the mentioned tensions\,\cite{Melchiorri2017,Valentino2017,SolGomCruzPLB2017,GomSolEPL2018}, and hence one would like to further elaborate  if the available observational data show a real preference for DDE over the $\CC$CDM.  In particular, the use of the large scale structure  (LSS) formation data seems to play an important role in the search for the DDE signal.  We find that by using the rich set of growth rate data  points $f(z)\sigma_8(z)$  presently available in the literature  the signal becomes crisper than if parameterized  in terms of the weak gravitational lensing data.  Since this fact is potentially very important for an eventual pinning down of the signal we illustrate it in the present study by using different models and parameterizations of the DE.

%%%%%%%%%%%%%%%%%%%%%%%%%%%%%%%%%%%%%%%%%%%%%%%%%%%%%%%%%%%%%%%%%%%%%%%%%%%%%%%%%%%%
%
\begin{figure*}
\centering
\includegraphics[angle=0,width=0.7\linewidth]{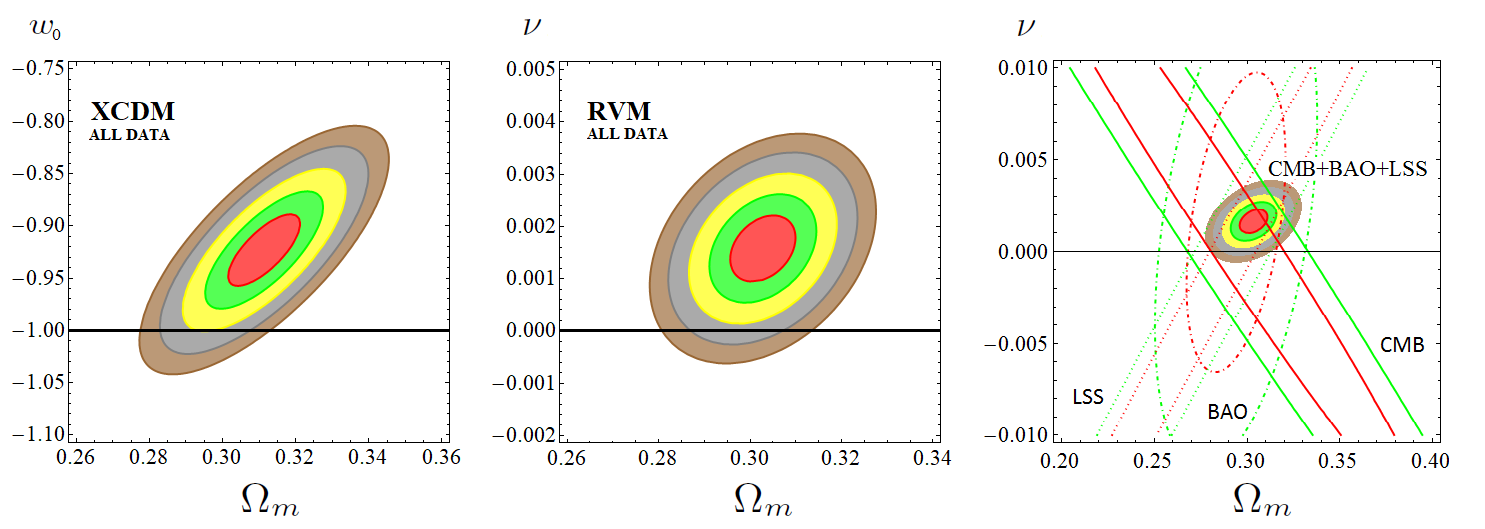}
\caption{\label{fig:XCDMEvolution}%
Likelihood contours from $1\sigma$ up to 5$\sigma$ c.l.  for the XCDM (left)  in the $(\Omega_m,w_0)$-plane, and the RVM (center)  in the $(\Omega_m,\nu)$-plane,  using all SNIa+BAO+$H(z)$+LSS+CMB  data after marginalizing over the rest of the fitting parameters indicated in Table 1. The plot on the right shows the contours for the RVM when only the CMB+BAO+LSS data are used. It also displays the partial contributions of these data sources at $1\sigma$ and $2\sigma$ . Further marginalization over $\Omega_m$ increases the c.l. of DDE up to $3.35\sigma$  (resp. $3.76\sigma$) for the XCDM (resp. RVM). The main contribution to the DDE signal is seen to emerge from the triad of  CMB+BAO+LSS data.
}
\end{figure*}
%
%%%%%%%%%%%%%%%%%%%%%%%%%%%%%%%%%%%%%%%%%%%%%%%%%%%%%%%%%%%%%%%%%%%%%%%%%%%%%%%%%%%
%%%%%%%%%%%%%%%%%%%%%%%%%%%%%%%%%%%%%%%%%%%%%%%%%%%%%%%%%%%%%%%%%%%%%%%%%%%%%%%%%%%%

%%%%%%%%%%%%%%%%%%%%%%%%%%%%%%%%%%%%%%%%%%%%%%%%%%%%%%%%%%%%%%%%%
%%%%%%%%%%%%%%%%%%%%%%%%%%%%%%%%%%%%%%%%%%%%%%%%%%%%%%%%%%%%%%%%%
%%%%%%%%%%%%%%%%%%%%%%%%%%%%%%%%%%%%%%%%%%%%%%%%%%%%%%%%%%%%%%%%%

%

\section{Dynamical dark energy}

Let us consider a generic cosmological framework described by the spatially flat Friedmann-Lema\^\i tre-Robertson-Walker (FLRW) metric, in which matter is in interaction with  a DDE density  $\rD(\zeta)$, which is a function of some dynamical variable $\zeta$ evolving with the cosmic time ($\dot{\zeta}\equiv d\zeta/dt\neq 0$).  Such variable can be e.g. the scale factor $a(t)$, the Hubble function $H(t)=\dot a(t)/a$ or some scalar field $\phi(t)$, all of them functions of the cosmic time. The corresponding pressure is given by $\pD=\wD\rD$, where $\wD$ is the equation of state (EoS) parameter of the DE. We assume $G=$const.  but $\dot{\rho}_{ D}\equiv (d\rho_{ D}/d\zeta) \dot\zeta\neq 0$.  In all these scenarios the Friedmann and acceleration equations with flat FLRW metric adopt the following generic form:
\begin{eqnarray}
3H^2&=&8\pi\,G\sum_N\rho_N\label{eq:FriedmannEq}\\
3H^2+2\dot{H}&=&-8\pi\,G\sum_Np_N\label{eq:PressureEq}\,,\,
\end{eqnarray}
where the sum is over all the components of the cosmic fluid: $N=dm,b,r,D$,  i.e. dark matter (DM), baryons, radiation and DE, with $w_N=p_N/\rho_N$ the EoS parameter for each component. The total nonrelativistic part reads
  $\rmm=\rb+\rdm$ and   involves the contributions from baryons and cold DM, both with vanishing pressure, whereas the radiation component  satisfies $p_r=\rho_r/3$.  In general the EoS of the DE can be a function of $\zeta$ and thus a nontrivial function of the cosmic time (e.g. when $\zeta$ is a scalar field with a given potential).  The particular case  $\wD=-1$ corresponds to the EoS for the vacuum energy $\rL$, whose pressure is given by $\pL=-\rL$.   The simplest possibility realizing this scenario corresponds to having constant vacuum energy density $\rL=\rLo=$const. and this defines the $\CC$CDM model. However, we will admit also the possibility of dynamical vacuum energy density, for which $\pL(t)=-\rL(t)$.
%Finally, in our case we assume that $\rL(\zeta)$ is a function of a cosmic variable $\zeta=\zeta(a)$, which ultimately depends on the scale factor or the cosmological redshift, $z=a^{-1}-1$.
The local conservation law associated to the above equations reads:
\begin{equation}\label{eq:GeneralCL}
\sum_N\dot{\rho}_{N}+3 H(1+w_N)\rho_N=0\,.
\end{equation}
If the DE part is conserved ($\dot{\rho}_{ D}+3 H(1+\wD)\rD=0$) the matter and radiation components are also conserved. This is e.g. the case when $\zeta$ is a self-conserved scalar field $\phi$ (see below).  In general, we may have an interacting scenario. For example, in dynamical models of the vacuum energy $\rD(t)=\rL(t)$ (with  $\wD=-1$ ) the last term of \eqref{eq:GeneralCL} vanishes and the cosmic evolution of $\rL$ is determined by some given source $Q$ which is model-dependent: $\dot{\rho}_\CC=-Q$, with  $0<|Q|\ll\dot{\rho}_m$ so as not to deviate too much from  the  $\CC$CDM model. For the latter, $Q=0$ since $\rL=$const. %In contrast to \,\cite{GomSolBas2015,ApJL2015,ApJ2017},
If we assume that radiation and baryons are self-conserved, so that their energy densities evolve in the standard way, i.e. $\rho_r(a)=\rho_{r0}\,a^{-4}$ and $\rho_b(a) = \rho_{b0}\,a^{-3}$, the possible dynamics of $\rL$ is exclusively associated to the exchange of energy with the DM. In this case, Eq.\,(\ref{eq:GeneralCL}) boils down to
\begin{equation}\label{eq:Qequations}
\dot{\rho}_{dm}+3H\rho_{dm}=Q\,,\ \ \ \ \ \ \, \dot\rho_{\CC}=-{Q}\,.
\end{equation}
In the literature, the source $Q$ is chosen {\it ad hoc} in a variety of forms -- see  e.g.\,\cite{Salvatelli2014,Murgia2016,Li2016,Costa2017}. Here we focus on a theoretically more appealing possibility, viz. the running vacuum model (RVM), which can be motivated in the context of QFT in curved spacetime and with possible implications for the physics of the early universe (see\,\cite{JSPRev2013,SolGom2015,Sola2015} and references therein). The cosmic variable $\zeta$ can then be identified not just with the cosmic time or the scale factor but with the full Hubble rate: $\zeta=H$. This special theoretical status of the RVM as compared to  other models considered in the literature is also phenomenologically advantageous.

%In the RVM the dynamical nature of the vacuum is governed by a renormalization group equation.

Well after inflation, the vacuum energy density in the RVM can be written in the relatively simple form \,\cite{JSPRev2013,SolGom2015,Sola2015}:
\begin{equation}\label{eq:RVMvacuumdadensity}
\rho_\CC(H) = \frac{3}{8\pi{G}}\left(c_{0} + \nu{H^2}\right)\,.
\end{equation}
The additive constant $c_0=H_0^2\left(\Omega_\CC-\nu\right)$ is fixed by the boundary condition $\rL(H_0)=\rLo$, where $\rLo$ and $H_0$ are the current values of these quantities, and $\OL$ is the present vacuum density parameter.
Theoretically, the dimensionless coefficient $\nu$ encodes the dynamics of the vacuum at low energy and can be related with the $\beta$-function of the running of $\rL$, so we naturally expect $|\nu|\ll1$. An estimate of $\nu$ in QFT indicates that it is of order $10^{-3}$ at most \cite{Fossil07}, but here we will treat it as a free parameter and hence we shall deal with the RVM on pure phenomenological grounds, thus fitting $\nu$ to the observational data.

In the RVM, the source function $Q$ in \eqref{eq:Qequations} is calculable from \eqref{eq:RVMvacuumdadensity} and Friedmann's equation (\ref{eq:FriedmannEq}), with the result $ Q=-\dot{\rho}_{\Lambda}=\nu\,H(3\rho_{m}+4\rho_r)$.
Thus, from the phenomenological point of view the dimensionless coefficient $\nu$ parameterizes both the evolution of the vacuum energy density and the strength of the dark-sector interaction, which in this way naturally satisfies the aforementioned condition  $0<|Q|\ll\dot{\rho}_m$. Furthermore, for $\nu>0$ the vacuum decays into DM (which is thermodynamically favorable) whereas for $\nu<0$ is the other way around.

The vacuum energy density can be computed straightforwardly from the above formulae. In the matter-dominated epoch, it behaves as
\begin{eqnarray}
\rho_\CC(a) &=& \rLo + \frac{\nu\,\rho_{m0}}{1-\nu}\left(a^{-3(1-\nu)}-1\right)\,. \label{eq:rhoVRVM}
\end{eqnarray}
Notice the correct normalization in our time $\rL(1)=\rLo$. Recall that the baryon and radiation densities adopt the standard  $\CC$CDM expressions as a function of the scale factor, as they are assumed not to interact with the vacuum. However, the CDM component does interact with it and therefore scales in an anomalous way.  As a result the energy density of the total matter component in our epoch scales as  $a^{-3(1-\nu)}$, which explains the possibility that the vacuum energy density can acquire a running part, see Eq.\, (\ref{eq:rhoVRVM}). When $\nu=0$ both the scaling of matter reduces to the standard one and the vacuum energy density (\ref{eq:rhoVRVM}) becomes constant as in the $\CC$CDM. The full Hubble function and energy densities include in general also the contribution from the radiation terms, which we have also taken into account, but we do not quote them explicitly here.

\subsection{XCDM and CPL parameterizations}   The simplest parameterization of the DDE is provided by the XCDM\cite{XCDM}: $\rho_D(a)=\rho_{D0}\,a^{-3(1+w_0)}$, with $\rho_{D0}=\rho_{\CC 0}$, where $w_D=w_0$ is the (constant) EoS parameter for  some unspecified DE entity X. For $w_0=-1$ it reduces to the CC term. The XCDM is useful to roughly mimic a (non-interactive) DE scalar field $\phi$ with $w_{\phi}\simeq -1$.  Our XCDM fit yields $w_0=-0.923\pm0.023$ (cf. Table 1), clearly pointing to quintessence behavior at $\sim 3.3\sigma$ c.l.   Fitting the same data to  the CPL parameterization\,\cite{CPL} with slowly evolving EoS,
%\begin{equation}\label{eq:CPL}
$w(a)=w_1+w_2\,(1-a)$, renders $w_1=-0.944\pm0.089$ and $w_2=0.063\pm0.259$.  As we can see the error on $w_1$, and specially on  $w_2$, are much larger than that of $w_0$  in the XCDM case owing to the extra parameter.  In what follows we shall not consider the CPL parameterization any longer. A more realistic quintessence model with dynamical EoS is considered next.
%\end{equation}

%%%%%%%%%%%%%%%%%%%%%%%%%%%%%%%%%%%%%%%%%%%%%%%%%%%%%%%%%%%%%%%%%%%%%%%%%%%%%%%%%%%
%%%%%%%%%%%%%%%%%%%%%%%%%%%%%%%%%%%%%%%%%%%%%%%%%%%%%%%%%%%%%%%%%%%%%%%%%%%%%%%%%%%%
\begin{figure*}
\centering
\includegraphics[angle=0,width=0.7\linewidth]{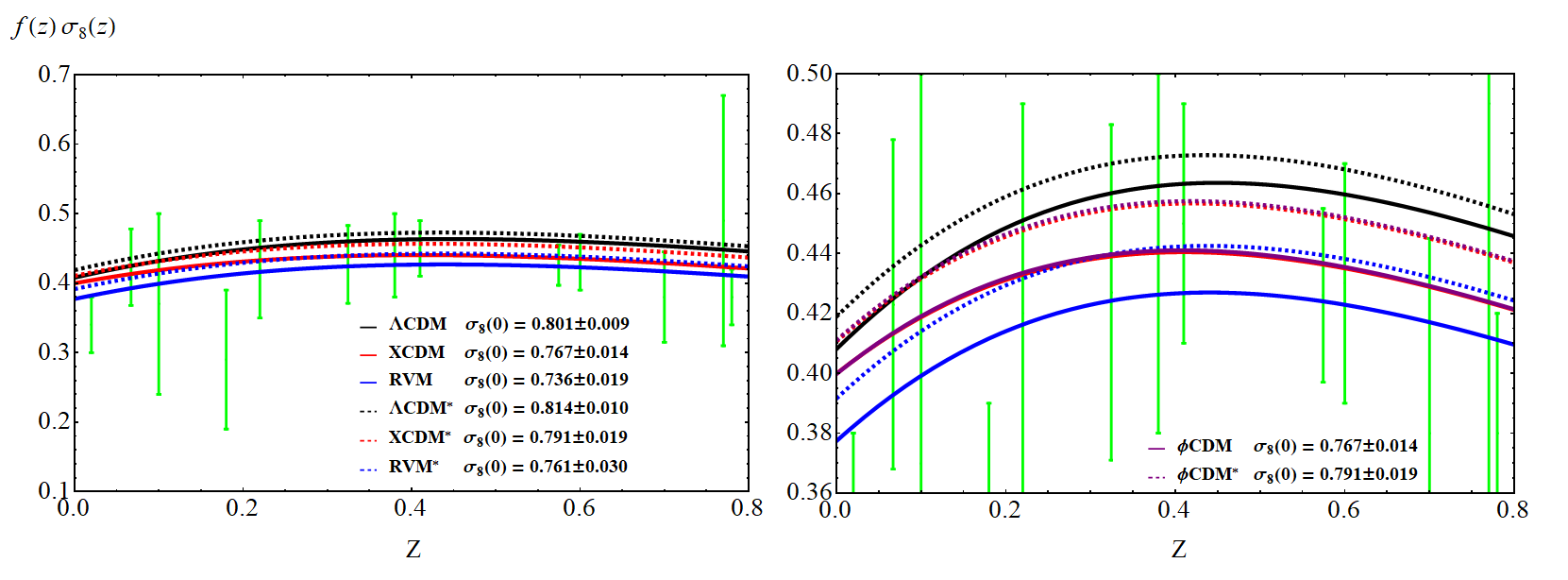}
\caption{\label{RecRVMtriad}%
The prediction of the various models confronted to the LSS data points $f(z)\sigma_8(z)$ for the normal and starred scenarios of Table 1. The plot on the right shows a magnified view and includes the $\phi$CDM prediction as well, which almost overlaps with that of the XCDM. The EoS analysis presented in Fig. 3 explains the possible origin of the large overlap (see also the text).}
\end{figure*}

%%%%%%%%%%%%%%%%%%%%%%%%%%%%%%%%%%%%%%%%%%%%%%%%%%%%%%%%%%%%%%%%%%%%%%%%%%%%%%%%%%
%%%%%%%%%%%%%%%%%%%%%%%%%%%%%%%%%%%%%%%%%%%%%%%%%%%%%%%%%%%%%%%%%
%%%%%%%%%%%%%%%%%%%%%%%%%%%%%%%%%%%%%%%%%%%%%%%%%%%%%%%%%%%%%%%%%
%

\subsection{$\phi$CDM models}

A natural question that can be formulated is whether the traditional class of $\phi$CDM models\,\cite{CCP}, in which the DE is described in terms of a scalar field $\phi$ with some standard form for its potential $V(\phi)$, are also capable of capturing consistent signs of DDE using the same set of cosmological observations. Remarkably enough, the answer is affirmative -- see Table 2. The traditional quintessence and phantom scalar fields \cite{PeeblesRatra88,Wetterich88,Caldwell98,ZlatevWangSteinhardt99,Amendola2000}  mentioned in the introduction are comprised within the $\phi$CDM class. If we take the scalar field $\phi$ dimensionless, its energy density and pressure are given by
\begin{equation}\label{eq:rhophi}
\rho_\phi=\frac{M^2_{P}}{16\pi}\left[\frac{\dot{\phi}^2}{2}+V(\phi)\right]\,,\ p_\phi=\frac{M^2_{P}}{16\pi}\left[\frac{\dot{\phi}^2}{2}-V(\phi)\right]\,.
\end{equation}
Here $M_P=1/\sqrt{G}=1.2211\times 10^{19}$ GeV is the  Planck mass (in natural units).
As a representative potential we borrow the original Peebles \& Ratra (PR) quintessence potential\,\cite{PeeblesRatra88}:
\begin{equation}\label{eq:PRpotential}
V(\phi)=\frac{1}{2}\kappa M_P^2\phi^{-\alpha}\,.
\end{equation}
For the motivation of this potential, see \cite{PeeblesRatra88}. Let us only recall that it admits tracker solutions of the field equations for $\alpha>0$. Parameters $\kappa$ and $\alpha$ are determined from our fit to the  cosmological data.
For the $\phi$CDM it is more convenient to use the free parameters in Table 2 and then compute $h$ and $\Omega_m$ from them.   Recalling that $H_0\equiv 100h\,\varsigma$, with $\varsigma\equiv 1 Km/s/Mpc=2.133\times10^{-44} GeV$ (in natural units), we define $\bar{\kappa}\equiv \kappa\,M_P^2/\varsigma^2$ and express the fitting resuts in terms of $\alpha$ and $\bar{\kappa}$\,\cite{MPLA2017}.  Using the best-fit values from Table 2 and the overall covariance matrix derived from our fit, we obtain  $h=0.671\pm 0.006$ and $\Omega_m=0.312\pm 0.006$, which can be compared with the corresponding values for the other models in Table 1.

The plot for the dynamical EoS of the $\phi$CDM, $w_{\phi}(z)=p_{\phi}(z)/\rho_{\phi}(z)$,  in terms of the redshift near our time is shown in Fig.\,3, together with the (constant) EoS value of the XCDM parametrization, including the $1\sigma$ error band in both cases. However, because of the dynamical character of $w_{\phi}(z)$ the central curve cannot be obtained as a direct output of the $\chi^2$ minimization procedure. We used a Monte Carlo analysis with the Metropolis-Hastings algorithm \cite{MetropolisHastings} in order to sample the exact distributions of $w(z_i)$, and obtained a Markov chain for each redshift $z_i$ from which to compute the mean and the associated standard deviation.
The behavior of the curves shows that the quintessence-like behavior is sustained until the present epoch. Note from Fig.\,3 that the central value of the constant EoS for the XCDM parameterization remains inside the $1\sigma$ region of the $\phi$CDM model in the relevant redshift range. Incidentally, this is the range in which the DE density starts to be significant and even dominant for $z\lesssim0.7$. This fact might explain why the simple XCDM parametrization can mimic so well the more complex and physically motivated $\phi$CDM model, and why the fitting performance and the description of the LSS data are alike in both scenarios (cf. Tables 1 and 2, and Fig. 2).
The likelihood contours for the PR model in the $(\Omega_m,\alpha)$-plane are depicted in Fig.\,4 (left). They clearly point to a nonvanishing and positive value of $\alpha$ at $\gtrsim3\sigma$ c.l. when all data are used (more discussion on these contours later on).
Bearing also in mind the  numerically computed EoS value at present ($z=0$), namely
\begin{equation}\label{eq:wphinow}
w_\phi(z=0)=-0.936\pm0.019\,,
\end{equation}
we learn  that the EoS parameter lies  $3.37\sigma$  away  from $-1$ into the quintessence region.  It is comparable to the result obtained from the XCDM parametrization in Table 1, which is $\sim 3.35\sigma$ above $-1$.
%%%%%%%%%%%%%%%%%%%%%%%%%%%%%%%%%%%%%%%%%%%%%%%%%%%%%%%%%%%%%%%%%%
\begin{table*}
\begin{center}
\begin{scriptsize}
\resizebox{1\textwidth}{!}{
\begin{tabular}{| c | c |c | c | c | c | c | c | c | c|c|}
\multicolumn{1}{c}{Model} &  \multicolumn{1}{c}{$\omega_m=\Omega_m h^2$} &  \multicolumn{1}{c}{$\omega_b=\Omega_b h^2$} & \multicolumn{1}{c}{{$n_s$}}  &  \multicolumn{1}{c}{$\alpha$} &  \multicolumn{1}{c}{$\bar{\kappa}(\times 10^{3})$}&  \multicolumn{1}{c}{$\chi^2_{\rm min}/dof$} &  \multicolumn{1}{c}{$\ln A$} & \multicolumn{1}{c}{$\ln B$}\vspace{0.5mm}
\\\hline
$\phi$CDM  &  $0.1405\pm 0.0008$& $0.02263\pm 0.00014 $&$0.976\pm 0.004$& $0.202\pm 0.065$  & $32.7\pm1.2$ &  74.08/84 & 4.28 & 3.16 \\
\hline
$\phi$CDM$^{*}$   &  $0.1416\pm 0.0010$& $0.02253\pm 0.00015 $&$0.973\pm 0.004$& $0.107\pm 0.086$  & $34.3\pm 2.5$ &  64.37/72 & -0.36 & -1.39 \\
\hline
 \end{tabular}
 }
\caption{The best-fit values for the $\phi$CDM model with PR potential (\ref{eq:PRpotential}). We use the same cosmological data set as for the other models in Table 1.  Apart from the standard parameters, we  have the specific model parameters $\alpha$  and  $\bar{\kappa}$ (see the text).  We find $\gtrsim 3\sigma$ c.l. evidence in favor of $\alpha>0$.  In terms of the EoS of $\phi$ at present, the DE behavior appears quintessence-like, see Eq.\,(\ref{eq:wphinow}). In Fig. 4 we show the contour lines of the model together with those of the RVM when we take different data sets.}
 \end{scriptsize}
\end{center}
\label{tableFitPhiCDM}
\end{table*}

%%%%%%%%%%%%%%%%%%%%%%%%%%%%%%%%%%%%%%%%%%%%%%%%%%%%%%%%%%%%%%%%%

%%%%%%%%%%%%%%%%%%%%%%%%%%%%%%%%%%%%%%%%%%%%%%%%%%%%%%%%%%%%%%%%%
%%%%%%%%%%%%%%%%%%%%%%%%%%%%%%%%%%%%%%%%%%%%%%%%%%%%%%%%%%%%%%%%%

\section{Dynamical DE and structure formation}
The analysis of the LSS data plays a crucial role and deserves some remarks. In the presence of dynamical vacuum the matter density contrast $\delta_m=\delta\rho_m/\rho_m$ obeys the following differential equation with respect to the scale factor\,\cite{GomSolBas2015}:
\begin{eqnarray}\label{diffeqDa}
&&{\delta}''_m(a)+\left[\frac{3}{a}+\frac{H'(a)}{H(a)}+\frac{\Psi(a)}{aH(a)}\right]\,{\delta}'_m(a)\phantom{XXXX}\nonumber\\
&&-\left[\frac{4\pi G\rmr(a)}{H^2(a)}
-\frac{2\Psi(a)}{H(a)}-a\frac{\Psi'(a)}{H(a)}\right]\,\frac{\delta_m(a)}{a^2}=0\,,\phantom{XXXX}
\end{eqnarray}
where $\Psi\equiv Q/{\rmr}$, and $Q$ is the interaction source, which as we have seen is known in the RVM case.  For $\rL=$const. and also for the XCDM and CPL, $Q=0$, and Eq.\,(\ref{diffeqDa})  reduces to the usual $\CC$CDM form. The same standard perturbations equation can be licitly used for the PR model\,\cite{PeeblesRatra88}, but using of course the corresponding Hubble function  $H^2=(8\pi G/3)(\rmm+\rho_{\phi})$. This is because  $\rmm$ and $\rho_{\phi}$ do not interact in this case  and hence  $\rho_m\sim  a^{-3}$.

To solve Eq.\,(\ref{diffeqDa}) numerically for the RVM we have to fix the initial conditions at high redshift.  One can check that the growing mode solution of  (\ref{diffeqDa}) in the limit of small scale factor into the matter-dominated era is given by the power-law  $\delta_m\sim a^{1-3\nu}$ (becoming $\delta_m\sim a$ for $\nu=0$, as expected). Using this form to fix the initial conditions for $\delta_m$ and $\delta'_m$ at large redshift, say at $z_{ini}\sim100$ ($a_{ini}\sim10^{-2}$), we can numerically solve for any other redshift down to the current value $z=0$, and the result does not depend significantly on $z_{ini}$ provided it is large enough but still well below decoupling ($z\sim 10^3$), where the radiation component starts to be significant. As for the $\phi$CDM case,  the Klein-Gordon equation with PR potential (\ref{eq:PRpotential}) in the flat FLRW metric in terms of the scale factor reads
\begin{equation}\label{eq:KGa}
\phi^{\prime\prime}(a)+\left(\frac{{H}^\prime(a)}{{H(a)}}+\frac{4}{a}\right)\phi^\prime(a)-
\frac{\alpha}{2}\frac{{\kappa}(\phi(a))^{-(\alpha+1)}}{(a{H(a)})^2}=0\,.
\end{equation}
At high redshift into the matter-dominated epoch, when $\rho_{\phi}$ can be neglected,  the Hubble function takes on the simple form  $H^2=(8\pi G/3) \rmo a^{-3}$. This allows to search once more for power-law solutions, and one finds  $\phi(a)\propto a^{3/(2+\alpha)}$ and the corresponding $\phi'(a)$. With these initials conditions, we can now find the exact solution of (\ref{eq:KGa}) for any value of the scale factor by numerical integration\,\cite{MPLA2017}.
%%%%%%%%%%%%%%%%%%%%%%%%%%%%%%%%%%%%%%%%%%%%%%%%%%%%%%%%%%%%%%%%%%%%%%%%%%%%%%%%%%%
%%%%%%%%%%%%%%%%%%%%%%%%%%%%%%%%%%%%%%%%%%%%%%%%%%%%%%%%%%%%%%%%%%%%%%%%%%%%%%%%%%%%
\begin{figure}
\centering
\includegraphics[angle=0,width=0.7\linewidth]{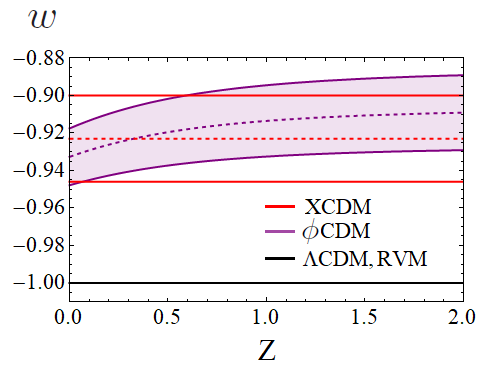}
\caption{\label{RecRVMtriad2}%
The EoS $w = w(z)$ for the XCDM and  $\phi$CDM models within the corresponding $1\sigma$ bands. For the XCDM the EoS is of course ``flat'' (constant): $w = -0.923\pm 0.023$  (cf. Table 1) and points to quintessence (at $3.35\sigma$ c.l.) As for the $\phi$CDM model, with  PR potential (\ref{eq:PRpotential}), the EoS evolves with time and is computed through a Monte Carlo analysis (see text). The current value reads as in Eq.\,(\ref{eq:wphinow}), which favors once more the quintessence region (at $3.37\sigma$ c.l.)}
\end{figure}

%%%%%%%%%%%%%%%%%%%%%%%%%%%%%%%%%%%%%%%%%%%%%%%%%%%%%%%%%%%%%%%%%%%%%%%%%%%%%%%%%%
%%%%%%%%%%%%%%%%%%%%%%%%%%%%%%%%%%%%%%%%%%%%%%%%%%%%%%%%%%%%%%%%%
%
\begin{figure*}
\centering
\includegraphics[angle=0,width=0.6\linewidth]{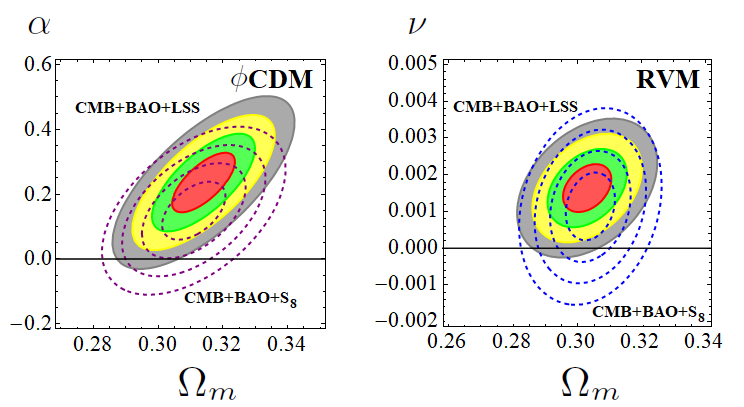}
\caption{\label{fig:fsigma8}
Contour lines for the $\phi$CDM with PR potential (\ref{eq:PRpotential}) (left) and RVM (\ref{eq:RVMvacuumdadensity}) (right) using the same CMB+BAO+LSS data as in Table 1 (solid contours); and also when replacing the LSS data (i.e. the $f(z)\sigma_8(z)$ points)  with the $S_8$ value obtained from the analysis of the weak gravitational lensing data\,\cite{Joudaki2017} (dashed lines), i.e. the starred scenarios in Tables 1 and 2.}
\end{figure*}
%%%%%%%%%%%%%%%%%%%%%%%%%%%%%%%%%%%%%%%%%%%%%%%%%%%%%%%%%%%%%%%%%%%%%%%%%%%%%%%%%
%%%%%%%%%%%%%%%%%%%%%%%%%%%%%%%%%%%%%%%%%%%%%%%%%%%%%%%%%%%%%%%%%%%%%%%%%%%%%%%%%%
The analysis of the linear LSS regime is performed with the help of the weighted linear growth $f(z)\sigma_8(z)$, where $f(z)=d\ln{\delta_m}/d\ln{a}$ is the usual growth factor and $\sigma_8(z)$ is the rms mass fluctuation amplitude on scales of $R_8=8\,h^{-1}$ Mpc at redshift $z$. Such amplitude reads \cite{ApJ2017}:
\begin{equation}
\begin{small}\sigma_{\rm 8}(z)=\sigma_{8, \CC}
\frac{\delta_m(z)}{\delta_{m}^\CC(0)}
\sqrt{\frac{\int_{0}^{\infty} k^{n_s+2} T^{2}(k,\vec{q})
W^2(kR_{8})\,dk} {\int_{0}^{\infty} k^{n_{s,\CC}+2} T^{2}(k,\vec{q}_\Lambda) W^2(kR_{8,\Lambda})\,dk}}\,,\label{s88general}
\end{small}\end{equation}
where $W$ is a top-hat smoothing function. Apart from the spectral index, $n_s$, the remaining fitting parameters in Table 1 are collected in the vector $\vec{q}$ involved in the transfer function  $T(k,\vec{q})$\,\cite{Bardeen}. Similarly, $n_{s,\CC}$ and $\vec{q}_\CC$ stand for the fixed parameters of the fiducial model, which we use to define the normalization of the power spectrum. As in \cite{ApJ2017} we take  the $\CC$CDM at fixed parameter values from the Planck 2015 TT,TE,EE+lowP+lensing analysis\,\cite{Planck2015}.

\section{Fitting results and discussion}

For the model comparison, we have defined a joint likelihood function ${\cal L}$. Assuming Gaussian errors, the total $\chi^2$ to be minimized reads:
\be
\chi^2_{tot}=\chi^2_{SNIa}+\chi^2_{BAO}+\chi^2_{H}+\chi^2_{LSS}+\chi^2_{CMB}\,.
\ee
Each one of these terms is defined in the standard way from the data\,\cite{DEBook} including the covariance matrices\,\cite{ApJ2017,PRD2017}.  The overall fit value of $\chi^2_{\rm min}$ for the DDE models is smaller than the $\CC$CDM one (cf. Tables 1 and 2).
For a better assessment of the situation, it proves useful to invoke the Akaike and Bayesian information criteria, AIC and BIC\,\cite{KassRaftery1995}:
%These efficient estimators are defined as
${\rm AIC}=\chi^2_{\rm min}+2nN/(N-n-1)$ and ${\rm BIC}=\chi^2_{\rm min}+n\,\ln N$, where $n$ is the number of fitting parameters and $N$ the number of data points.
The differences $\Delta$AIC ($\Delta$BIC) are computed with respect to the model that carries smaller value of AIC (BIC) -- e.g. the XCDM, $\phi$CDM and RVM here. In Tables 1 and 2 we quote $\ln A\equiv \Delta{\rm AIC}/2$ and $\ln B\equiv \Delta{\rm BIC}/2$. They provide the Akaike and Bayesian evidences (e.g. the Bayes factor $B$ yields the ratio of marginal likelihoods between the two models\,\cite{DEBook}).
In our context, for values of  $\ln A$ and $\ln B$  above $+3\, (+5)$ we are entitled to speak of ``strong'' (to ``very strong'') evidence in detriment of the $\CC$CDM and hence in support  of DDE\,\cite{KassRaftery1995}. In this language, Tables 1 and 2 denote strong (moderate) evidence for the XCDM and $\phi$CDM, but very strong evidence for the RVM (for which $\ln A, \ln B$ are both above $+5$).

Figures  1 and 4 display the contour plots for the models, providing also some details on the impact of different combination of data sets. For instance, in Fig.\,1 (right) it is seen that the CMB+BAO+LSS combination is crucial to pin down the DDE signature. The importance of the LSS data is further emphasized in  Fig. 2, where we display  $f(z)\sigma_8(z)$ for the various models. Only the DDE models can provide lower enough LSS power as to correctly match the data points, in stark contrast to the $\CC$CDM. We observe that the XCDM and $\phi$CDM curves in that figure (being almost overlapping) have less $f(z)\sigma_8(z)$ power than the $\CC$CDM curve and for this reason they fit the data points better. This is specially so for the RVM, which has even lesser LSS power and therefore further improves the fit quality of these data, what ultimately has a significant impact on the overall quality of the fit. The RVM has indeed the capacity to fully resolve the $\sigma_8$-tension mentioned in the introduction, as has recently been proven in \cite{GomSolEPL2018}.
Overall, the RVM  captures a crisp signal of DDE near $3.8\sigma$ while the XCDM and the $\phi$CDM (with PR potential) furnish a mutually consistent signal of more than $3.3\sigma$.  Remarkably, similar support to DDE (at $3.5\sigma$ c.l.) was recently reported in\,\cite{GongBoZhao2017} using nonparametric methods.

Owing to the role played by the structure formation data we further inquire into its impact when we use a different proxy to describe them.  The starred scenarios in Tables 1-2 and Fig. 4 explore the reaction of the fit when we replace the LSS data points  $f(z)\sigma_8(z)$  with the measurements  from the weak gravitational lensing, as it is done in many works in the literature. The weak-lensing data are usually encoded in terms of the effective parameter
 $
 S_8\equiv \sigma_8 \left(\Omega_m/{0.3}\right)^{0.5}
 $ (see e.g. \cite{Joudaki2017,Henning2017,Hildebrandt2017,Heymans2013}).
 For definiteness we use the recent study  by\cite{Joudaki2017},  in which they carry a combined analysis of cosmic shear tomography, galaxy-galaxy lensing tomography, and redshift-space multipole power spectra using imaging data by the Kilo Degree Survey (KiDS-450) overlapping with the 2-degree Field Lensing Survey  (2dFLenS) and the Baryon Oscillation Spectroscopic
Survey (BOSS).  They find $S_8=0.742\pm 0.035$.  {Incidentally, this  value is $2.6\sigma$ below the one provided by Planck's TT+lowP analysis \cite{Planck2015}}.  Our  conclusions  remain essentially unchanged if we use e.g. the $S_8$-constraints from \cite{Henning2017,Hildebrandt2017,Heymans2013}.
 The outcome of this additional check therefore shows that the use of  the weak-lensing data from  $S_8$ as a replacement for the direct LSS  measurements is insufficient since it definitely weakens the evidence in favor of DDE;  namely, the values of  $\ln A$  and $\ln B$  become smaller and even negative  in some cases (cf. the starred scenarios in Tables 1 and 2). From Fig. 4 we further confirm (using both the $\phi$CDM with  PR potential and the RVM) that the contour lines computed from the data string  CMB+BAO+LSS are mostly contained within the contour lines from the alternative string CMB+BAO+$S_8$ and are shifted upwards. The former data set is therefore more precise and capable of resolving the DDE signal at a level of more than $3\sigma$, whereas with $S_8$ it barely surpasses the  $1\sigma$ c.l., thus rendering the signal uncertain.

 {It would be interesting to check  if the same updated data string used here could produce  a similar level of improvement with other models of the DE, such as e.g. modified gravity theories, which have a potential for improving the cosmographic description\,\cite{Capozziello2014}. Such analysis, however, is beyond the scope of the present Letter.}

\section{Conclusions}

We find that a  rigid $\CC$-term in Einstein's equations despite being the simplest hypothesis may well not be the most favored one at present, namely when we put it in hard-fought competition with dynamical dark energy (DDE) models confronted to a large set of cosmological data comprising such crucial ingredients as CMB+BAO+LSS. Consistent signs of DDE in between $3-4\sigma$ c.l. (strongly supported by information criteria) are found using different models.  We conclude that the current cosmological data do uphold in a significant fashion a mild evolution of the dark energy, in contrast to the concordance model with $\CC=$const.

We hope that our results might inject some more optimism for an eventual solution of the cosmological constant and coincidence problems.

\acknowledgments

\noindent We are partially supported by FPA2016-76005-C2-1-P, 2017-SGR-929 (Generalitat de
Catalunya) and  MDM-2014-0369 (ICCUB).


\begin{thebibliography}{}

%%%%%%%%%%%%%%%%%%%%%%%%%%%%%%%%%%%%%%%%%%%%%%%%%%%%%%%%%%%%%%%%%%%%%%%%%%%%%%%%%%%
%%%%%%%%%%%%%%%%%%%%%%%%%%%%%%%%%%%%%%%%%%%%%%%%%%%%%%%%%%%%%%%%%%%%%%%%%%%%%%%%%%%%


\bibitem{Einstein1917} Einstein A., \textit{Kosmologische Betrachtungen zur allgemeinen Relativit\"atstheorie},
Sitzungsber. K\"onigl. Preuss. Akad. Wiss. phys.-math. Klasse VI (1917)  142. %(submitted on February 8th 1917).

\bibitem{SNIaRiess}
 Riess A.G. {\it et al.},
  %``Observational evidence from supernovae for an accelerating universe and a cosmological constant,''
  Astron. J.,  {\bf 116}  (1998) 1009.

\bibitem{SNIaPerl}
  Perlmutter S. {\it et al.},
  %``Measurements of Omega and Lambda from 42 high redshift supernovae,''
  ApJ, {\bf 517} (1999) 565.	

\bibitem{Planck2015}
Ade P.A.R.  {\it et al.} (PLANCK Collaboration), A\&A, {\bf 594}  (2016) A13.

\bibitem{Weinberg}
Weinberg S., Rev. Mod. Phys., {\bf 61}  (1989) 1.

\bibitem{CCP}
Padmanabhan T., Phys. Rept., {\bf 380} (2003) 235;
Peebles P.J.E.  and Ratra B., Rev. Mod. Phys., {\bf 75} (2003)  559;  	
Copeland E.J., Sami M. and Tsujikawa S., Int. J. Mod. Phys., D{\bf 15} (2006) 1753.

\bibitem{DEBook} Amendola L. and Tsujikawa S., \textit{Dark Energy} (Cambridge Univ. Press, Cambridge, 2010 and 2015).

\bibitem{PSW}
Peccei R.D., Sol\`a J. and Wetterich C., Phys. Lett. B, \textbf{195} (1987) 183.


\bibitem{PeeblesRatra88}
Peebles P.J.E.  and Ratra B., ApJ Lett., {\bf 325} (1988)  L17; Phys. Rev., D\textbf{37} (1988) 3406.

\bibitem{Wetterich88}
Wetterich C., Nucl. Phys.,  B{\bf 302}  (1988) 668.
%Astron. Astrophys. {\bf 301}, 321 (1995).

\bibitem{Caldwell98}
Caldwell R.R., Dave  R.  and Steinhardt P.J.,
%\textit{Cosmological imprint of an energy component with general equation of state},
Phys. Rev. Lett., {\bf 80} (1998)  1582.
%[arXiv:astro-ph/9708069].

\bibitem{ZlatevWangSteinhardt99}
 Zlatev I.,  Wang L.M. and Steinhardt  P.J., Phys. Rev. Lett., \textbf{82} (1999)  896.
%Phys. Rev. D\textbf{59}, 123504 (1999).


%\bibitem{IvayloWangSteinhardt99}
% 	Quintessence, cosmic coincidence, and the cosmological constant
%Z. Ivaylo, L.M. Wang, \& P.J. Steinhardt, Phys. Rev. Lett. \textbf{82} (1999) 896.

\bibitem{Amendola2000}
Amendola L., 	
%Coupled quintessence
Phys. Rev. D{\bf 62} (2000)  043511.
%e-Print: astro-ph/9908023
%Phys.Rev. D69 (2004) 103524.
%astro-ph/0311175


\bibitem{Capozziello2014}
{Capozziello S.  {\it et al.}, % Farooq O., Luongo O. and Ratra B.,
Phys. Rev. D{\bf 90} (2014)  044016.}

\bibitem{Riess2016}
Riess A.G. {\it et al.}, ApJ., {\bf 826} (2016) 56.
%e-Print: arXiv:1604.01424

\bibitem {Macaulay2013}
%Lower Growth Rate from Recent Redshift Space Distortion Measurements than Expected from Planck
Macaulay E., Wehus I.K. and Eriksen H.C., Phys. Rev. Lett., {\bf 111} (2013) 161301.
%e-Print: arXiv:1303.6583

\bibitem{BasilakosNesseris2017}
{Basilakos S. and Nesseris S., Phys. Rev.,  D{\bf 96} (2017) 063517.}
%e-Print: arXiv:1705.08797


\bibitem{BetouleJLA}
Betoule M.  {\it et al.}, A\&A, {\bf 568}  (2014)  A22.

\bibitem{GilMarin2017}
Gil-Mar\'in  H. {\it et al.}, MNRAS, {\bf465} (2017)  1757.
%arXiv:1606.00439 v2

\bibitem{Beutler2011}
Beutler F. {\it et al.},
%Mon.Not.Roy.Astron.Soc.
MNRAS,  {\bf 416} (2011) 3017.

\bibitem{Ross2015}
Ross A.J.  {\it et al.},
%, L. Samushia, C. Howlett, W.J. Percival, A. Burden \& M. Manera, %Mon.Not.Roy.Astron.Soc.
MNRAS,  {\bf 449}  (2015)  835.

\bibitem{Kazin2014}
Kazin E.A.  {\it et al.},
%Mon.Not.Roy.Astron.Soc.
MNRAS,  {\bf 441} (2014)  3524.

\bibitem{Delubac2015}
Delubac T. {\it et al.}, A\&A, {\bf 574} (2015) A59.

\bibitem{Aubourg2015}
Aubourg E. {\it et al.}, Phys. Rev., D{\bf 92} (2015)  123516 .


\bibitem{Zhang2014}
 Zhang C. {\it et al.},
%H. Zhang, S. Yuan, T-J. Zhang \& Y-C. Sun,
Res. Astron. Astrophys., {\bf 14} (2014) 1221.

\bibitem{Jimenez2003}
Jim\'enez R. {\it et al.},
%L. Verde, T. Treu \& D. Stern,
ApJ., {\bf 593} (2003) 622.

\bibitem{Simon2005}
Simon J.,  Verde  L.  and  Jim\'enez R., Phys. Rev., D{\bf 71}  (2005) 123001.

\bibitem{Moresco2012}
Moresco M.  {\it et al.}, JCAP, {\bf 1208} (2012)  006.

\bibitem{Moresco2016}
Moresco M.  {\it et al.}, JCAP, {\bf 1605} (2016)  014.

\bibitem{Stern2010}
Stern D. {\it et al.},
%,R. Jim\'enez, L. Verde, M. Kamionkowski \& S.A. Stanford,
 JCAP, {\bf 1002} (2010)  008.
 %e-Print: arXiv:0907.3149

\bibitem{Moresco2015}
Moresco M.,
%Mon.Not.Roy.Astron.Soc.
MNRAS,  {\bf 450}  (2015) L16.

\bibitem{Beutler2012}
Beutler F. {\it et al.},
%Mon.Not.Roy.Astron.Soc.
MNRAS,  {\bf 423} (2012) 3430.

\bibitem{Feix2015}
Feix M., Nusser A.  and  Branchini E., Phys. Rev. Lett., {\bf 115}  (2015) 011301.

\bibitem{Simpson2016}
Simpson F.  {\it et al.}, Phys. Rev., D{\bf 93}  (2016)  023525.

\bibitem{Blake2013}
Blake C.  {\it et al.}, MNRAS,  {\bf 436} (2013) 3089.
%Mon.Not.Roy.Astron.Soc.

\bibitem{Blake2011BAO}
Blake C.  {\it et al.},
%Mon.Not.Roy.Astron.Soc.
MNRAS,  {\bf 415} (2011)  2876.

\bibitem{Springob2016}
Springob C.M.  {\it et al.},
%Mon.Not.Roy.Astron.Soc.
MNRAS,  {\bf 456} (2016)  1886.

\bibitem{Granett2015}
Granett B.R.  {\it et al.}, A\&A, {\bf 583} (2015)  A61.

\bibitem{Guzzo2008}
Guzzo L. {\it et al.}, Nature, {\bf 451}  (2008) 541.

\bibitem{SongandPercival2009}
Song Y-S.  and Percival W.J., JCAP, {\bf 0910}  (2009)  004.


\bibitem{Huang2015}
Huang Q.G., Wang K.  and Wang S., JCAP, {\bf 1512} (2015)  022.
%e-Print: arXiv:1509.00969

\bibitem{ApJ2017} Sol\`a J., G\'omez-Valent A. and de Cruz P\'erez J., ApJ, {\bf 836} (2017) 43.
%arXiv:1602.02103.

\bibitem{PRD2017}
Sol\`a J., de Cruz P\'erez J. and G\'omez-Valent A.,
%\textit{Towards the firsts compelling signs of vacuum dynamics in modern cosmological observations},
arXiv:1703.08218.

\bibitem{Joudaki2017}
{Joudaki S. {\it et al.}, MNRAS {\bf 474} (2018) 4894.}

\bibitem{Capozziello2017}
{Capozziello S.  {\it et al.}, %D'Agostino R. and Luongo O.,   	
%\textit{Cosmographic analysis with Chebyshev polynomials},
arXiv:1712.04380.}

\bibitem{GomAmen2018}
% 	H0 from cosmic chronometers and Type Ia supernovae, with Gaussian Processes and the novel Weighted Polynomial Regression method
{G\'omez-Valent A. and Amendola L., arXiv:1802.01505.}

\bibitem{ApJL2015}
Sol\`a J., G\'omez-Valent A. and de Cruz P\'erez J., ApJ Lett., {\bf 811} (2015) L14.

\bibitem{GongBoZhao2017}  Zhao G.B. {\it et al.},
%\textit{Dynamical dark energy in light of the latest observations},
Nat. Astron., {\bf 1} (2017)  627.
%arXiv:1701.08165.


\bibitem{Melchiorri2017}
Valentino E.D.  {\it et al.},
%Melchiorri  A., Linder E.V. and J. Silk,  %\textit{Constraining dark energy dynamics in extended parameter space}, %arXiv:1704.00762.
Phys. Rev., D{\bf 96} (2017)  023523.


\bibitem{Valentino2017}
Valentino E.D.,  Melchiorri A. and  Mena O., Phys. Rev., D{\bf 96}  (2017) 043503.
%Can interacting dark energy solve the H0 tension?}, arXiv:1704.08342

\bibitem{SolGomCruzPLB2017}
Sol\`a J., G\'omez-Valent A. and de Cruz P\'erez J., Phys. Lett. B, {\bf 774} (2017) 317;
 %arXiv:1705.06723.
 Int. J. Mod. Phys., A{\bf 32} (2017) 1730014.
%earXiv:1709.07451

\bibitem{GomSolEPL2018}
{G\'omez-Valent A. and Sol\`a  J., EPL {\bf 120} (2017) 39001.}

\bibitem{Salvatelli2014}
Salvatelli V. {\it et al.}, Phys. Rev. Lett., {\bf 113} (2014)  181301.
% arXiv:1406.7297

\bibitem{Murgia2016}
Murgia R., Gariazzo S. and  Fornengo N., JCAP, {\bf 1604} (2016)  014.

\bibitem{Li2016} % 	Testing models of vacuum energy interacting with cold dark matter
Li Y.H.,  Zhang J.F. and  Zhang X., Phys. Rev., D{\bf 93} (2016)  023002. %arXiv:1506.06349?
%Constraints on the Coupling between Dark Energy and Dark Matter from CMB data

\bibitem{Costa2017}  Costa A. A.  {\it et al.},
%Xu X.-D.,  Wang B.  and Abdalla E.,
 JCAP, {\bf 1701}  (2017) 028.
%e-Print: arXiv:1605.04138


\bibitem{JSPRev2013}
Sol\`a J.,
%\textit{Cosmological constant and vacuum energy: old and new ideas},
 J. Phys. Conf. Ser., {\bf 453} (2013) 012015.
%e-Print: arXiv:1306.1527.

\bibitem{SolGom2015}
Sol\`a J. and G\'omez-Valent A.,
%\textit{The $\bar{\CC}$CDM cosmology: From inflation to dark energy through running $\CC$},
\,Int. J. Mod. Phys., D{\bf 24} (2015) 1541003.

\bibitem{Sola2015}
Sol\`a J., %\The cosmological constant and entropy problems: mysteries of the present with profound roots in the past
%e-Print:
Int. J. Mod. Phys., D{\bf 24} (2015) 1544027.
%[e-Print: arXiv:1505.05863].

\bibitem{Fossil07}
Sol\`a J., J. of Phys.,  A{\bf 41} (2008) 164066.
%[arXiv:0710.4151].

\bibitem{XCDM}  Turner S.M. and  White M., Phys. Rev., D{\bf 56}  (1997) R4439.

    \bibitem{CPL}  Chevallier M. and Polarski  D., Int. J. Mod. Phys., D{\bf 10} (2001) 213;  Linder E.V., Phys. Rev. Lett., {\bf 90}  (2003) 091301.

        \bibitem{MPLA2017}
Sol\`{a} J.,  G\'omez-Valent A. and  de Cruz P\'erez J., Mod. Phys. Lett., A{\bf 32} (2017)  1750054.
%[arXiv:1610.08965].
%\textit{Dynamical dark energy: scalar fields and running vacuum}, arXiv:1610.08965

\bibitem{MetropolisHastings}
Metropolis N. {\it et al.}, J. Chem. Phys., {\bf 21}  (1953)  1087; Hastings W.K., Biometrika, {\bf 57} (1970) 97.
%A. Rosenbluth, M. Rosenbluth, A. Teller, and E. Teller, J. Chem. Phys. {\bf 21}, 1087 (1953).}

\bibitem{GomSolBas2015}
G\'omez-Valent A.,  Sol\`{a}  J. and Basilakos S., JCAP, {\bf 1501} (2015)  004;
Basilakos S. and  Sol\`a J.,  Phys.Rev., D{\bf 90} (2014) 023008;
%e-Print: arXiv:1402.6594
 Basilakos S., Plionis  M.  and Sol\`{a}  J.,  Phys. Rev., D{\bf 80}  (2009) 083511.
 %arXiv:0907.4555

\bibitem{Bardeen}
Bardeen J.M.  {\it et al.}, % Bond  J.R., Kaiser N. and  Szalay A.S.,
ApJ, {\bf 304} (1986)  15.

  \bibitem{KassRaftery1995}
Kass R.E. and Raftery A., J. Amer. Statist. Assoc., {\bf 90} (1995) 773.


\bibitem{Henning2017}
{Henning J.W. {\it et al.} (SPT Collab.), ApJ {\bf 852} (2018) 97.}

\bibitem{Hildebrandt2017}
Hildebrandt H. {\it et al.}, MNRAS, {\bf 465} (2017) 1454.
% arXiv:1606.05338

\bibitem{Heymans2013}
Heymans C. {\it et al.}, MNRAS, {\bf 432} (2013) 2433.
% arXiv:1303.1808







%================================================================================================






\end{thebibliography}
\end{document}